\documentclass[floatfix,a4paper,tightenlines,amsmath,amssymb,nofootinbib,showpacs,notitlepage,showkeys,prd,onecolumn,10pt,aps,superscriptaddress]{revtex4-2}

\usepackage[utf8]{inputenc}
\usepackage{amsmath}
\usepackage{amssymb}
\usepackage{graphicx}
\usepackage{hyperref}
\usepackage[export]{adjustbox}
\usepackage[update]{epstopdf}
\usepackage{multirow}
\usepackage{placeins}

\newcommand{\fref}[1]{Fig. \ref{#1}}
\newcommand{\tref}[1]{Tab.~\ref{#1}}

\pdfoutput=1

\allowdisplaybreaks 

\begin{document}

\title{Quenched glueballs in the DSE/BSE framework}

\thanks{Contribution to the proceedings of the \textit{19th International Conference on Hadron Spectroscopy and Structure in memoriam Simon Eidelman}, Mexico City, July 26-30, 2021}

\author{Markus Q.~Huber}
\email{markus.huber@physik.jlug.de}
\affiliation{Institut f\"ur Theoretische Physik, Justus-Liebig--Universit\"at Gie\ss en, 35392 Gie\ss en, Germany}

\author{Christian S. Fischer}
\email{christian.fischer@theo.physik.uni-giessen.de}
\affiliation{Institut f\"ur Theoretische Physik, Justus-Liebig--Universit\"at Gie\ss en, 35392 Gie\ss en, Germany}
\affiliation{Helmholtz Forschungsakademie Hessen f\"ur FAIR (HFHF), GSI Helmholtzzentrum f\"ur Schwerionenforschung, Campus Gie{\ss}en, 35392 Gie{\ss}en, Germany}

\author{H\`elios Sanchis-Alepuz}
\email{helios.sanchis-alepuz@silicon-austria.com}
\affiliation{Silicon Austria Labs GmbH, Inffeldgasse 33, 8010 Graz, Austria}

\date{\today}

\begin{abstract}
The spectrum of glueballs with quantum numbers $J^\mathsf{PC}=0^{\pm+},2^{\pm+},3^{\pm+},4^{\pm+}$ is calculated in quenched quantum chromodynamics from bound state equations.
The input is taken from a parameter-free calculation of two- and three-point functions.
Our results agree well with lattice results where available and contain also some additional states.
For the scalar glueball, we present first results for the effects of additional diagrams which turn out to be strongly suppressed.
\end{abstract}

\maketitle

\section{Introduction}

The determination of the spectrum of bound states consisting dominantly of gluonic constituents is challenging both for theory and experiment, see, e.g., \cite{Klempt:2007cp,Crede:2008vw,Mathieu:2008me,Ochs:2013gi,Llanes-Estrada:2021evz}.
Prominent candidates for scalar and tensor glueballs are the measured $f_0$ and $f_2$ states.
Their gluonic content as determined from radiative $J/\psi$ decays was discussed recently in \cite{Sarantsev:2021ein,Rodas:2021tyb,Klempt:2021wpg}.
On the theory side, the benchmark has been lattice simulations of the glueball spectrum \cite{Bali:1993fb,Morningstar:1999rf,Chen:2005mg,Athenodorou:2020ani}.
However, the most reliable calculations were done for quenched quantum chromodynamics (QCD), viz., with all quark contributions suppressed.
The inclusion of dynamical quarks remains a challenge, see, e.g., \cite{Gregory:2012hu,Brett:2019tzr,Chen:2021dvn}.

Functional methods are an alternative nonperturbative method that can be used to calculate the hadron spectrum.
The determination of mesons and baryons is an active field, see, e.g., \cite{Cloet:2013jya,Eichmann:2016yit}.
Often, an effective interaction is employed in such calculations.
Such an approach was also employed for glueballs \cite{Meyers:2012ka,Sanchis-Alepuz:2015hma,Souza:2019ylx,Kaptari:2020qlt}.
Alternatively, one can determine the input in a self-contained way and eliminate the need for any model parameters.
This was achieved for the first time in \cite{Huber:2020ngt} where we observed that the self-consistency of the employed input is crucial.
The original work for spin $J=0$ was extended to higher spin in \cite{Huber:2021yfy}.
An extension of this setup for the pseudoscalar glueball was presented in \cite{Huber:2021zqk} where the effect of further diagrams was investigated and found to be subleading.
Here we study the same effect for the scalar glueball.
Besides the calculation of the spectrum from bound state equations, other functional approaches exist like the calculation from correlation functions of gauge invariant operators based on, e.g., an infrared momentum analysis \cite{Dudal:2010cd,Dudal:2013wja} or a direct calculation \cite{Windisch:2012sz}.
Also Hamiltonian many body methods \cite{Szczepaniak:1995cw,Szczepaniak:2003mr} or chiral Lagrangians
\cite{Janowski:2011gt,Eshraim:2012jv} can be used.

In this contribution we review the results for the quenched glueball spectrum \cite{Huber:2020ngt,Huber:2021yfy,Huber:2021zqk} and present first results taking into account additional diagrams for the scalar glueball.
We will first present the functional equations used to calculate the glueball spectrum and discuss how to solve them in Sec.~\ref{sec:methodology}.
The results are shown in Sec.~\ref{sec:results} which is followed by a summary.

\section{Methodology}
\label{sec:methodology}

\begin{figure*}[tb]
	\includegraphics[width=0.98\textwidth]{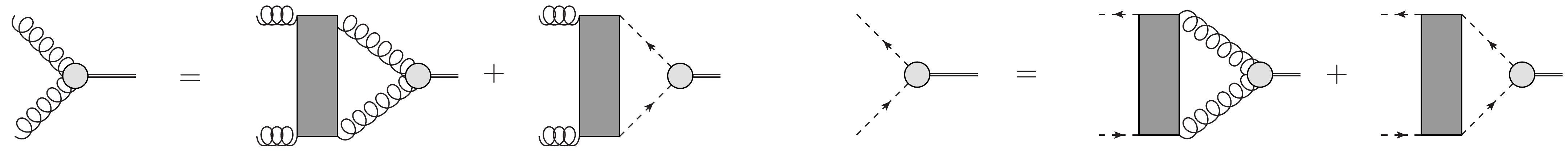}
	\caption{
		The coupled set of BSEs for a glueball made from two gluons and a pair of Faddeev-Popov (anti-)ghosts.
		Wiggly lines denote dressed gluon propagators, dashed lines denote dressed ghost propagators. 
		The gray boxes represent interaction kernels given in Fig.~\ref{fig:kernels}. The Bethe-Salpeter amplitudes of the glueball
		are denoted by gray disks. \label{fig:bses}
	}
	\vspace*{6mm}
	\includegraphics[width=0.48\textwidth]{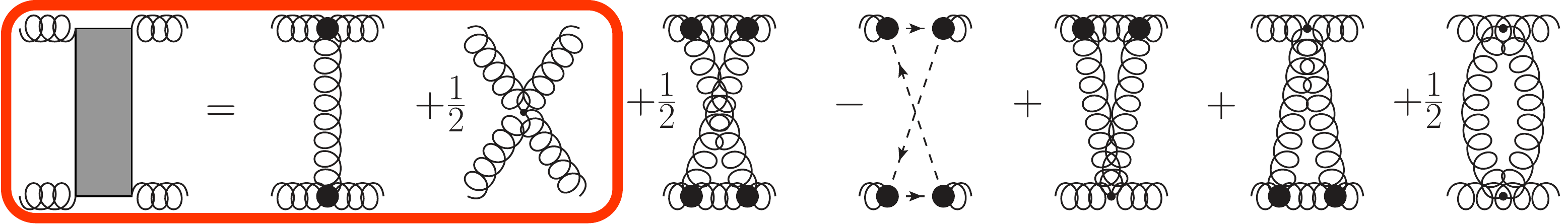}\\ 
	\vskip4mm
	\includegraphics[height=1.2cm]{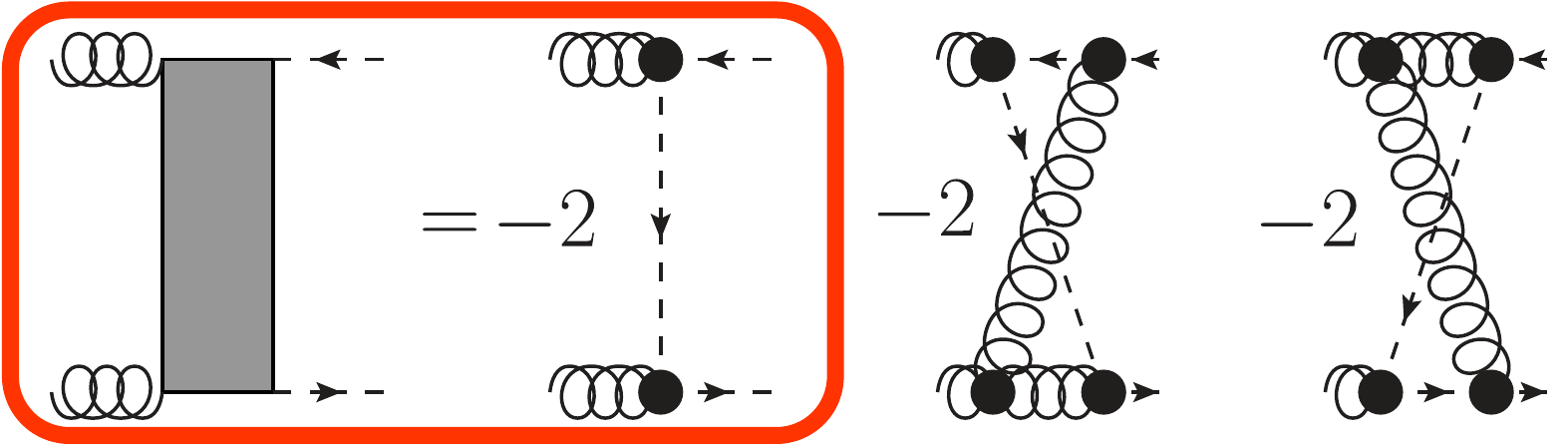}\hfill
	\includegraphics[height=1.2cm]{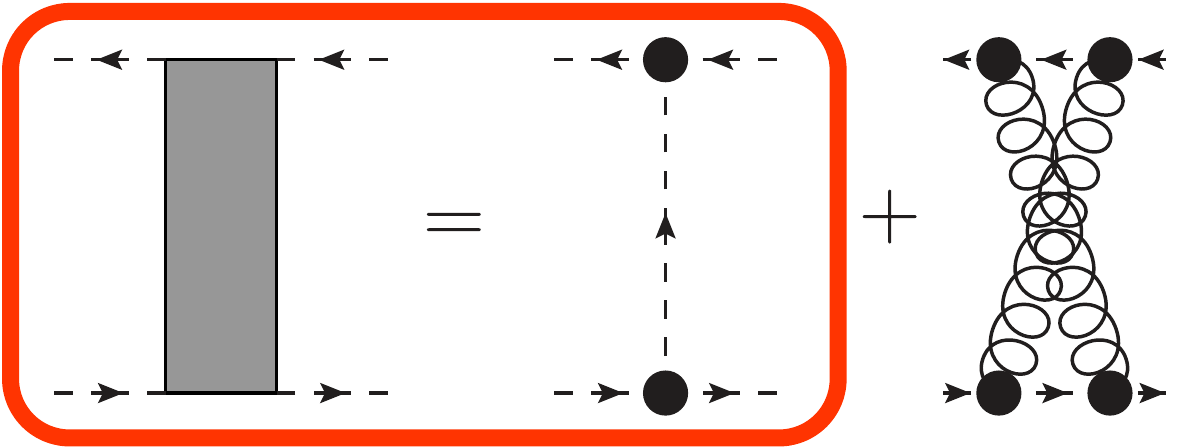}\hfill
	\includegraphics[height=1.2cm]{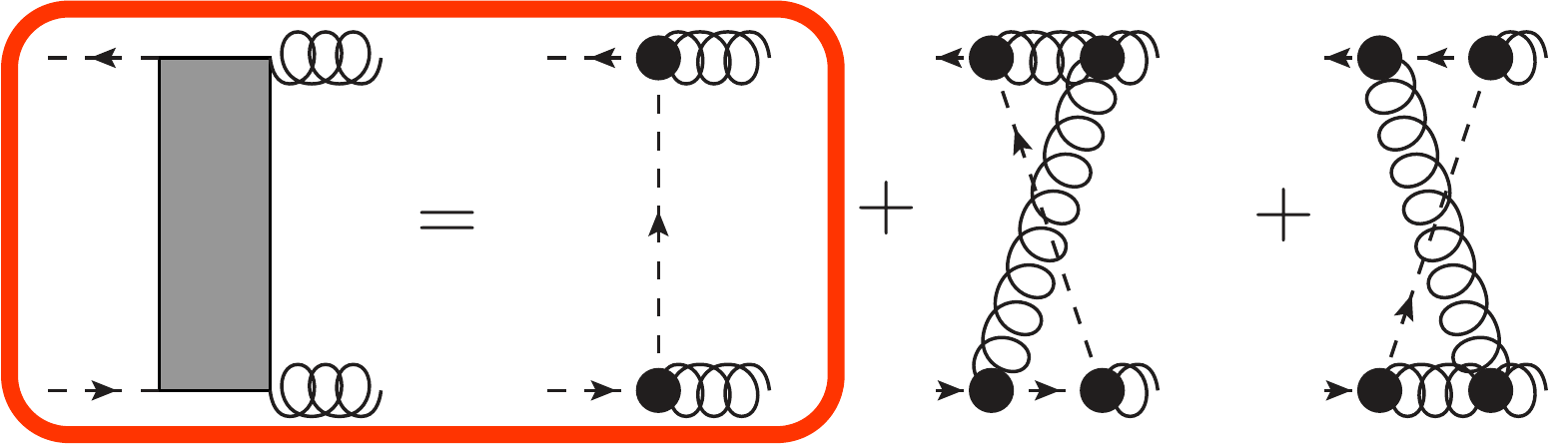}
	\caption{
		Interaction kernels from the three-loop 3PI effective action.
		All propagators are dressed; black disks represent dressed vertices.
		In our main calculation, we include the diagrams inside the red rectangles.
		For the extended calculation, all diagrams of the gluon-gluon interaction kernel (first line) are included.
		\label{fig:kernels}
	}
\end{figure*}

\begin{figure*}[t]
	\includegraphics[width=0.48\textwidth]{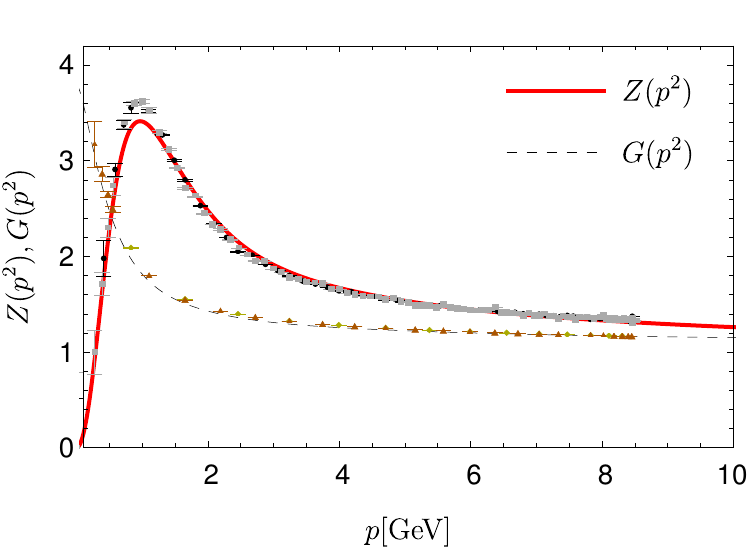}\hfill
	\includegraphics[width=0.48\textwidth]{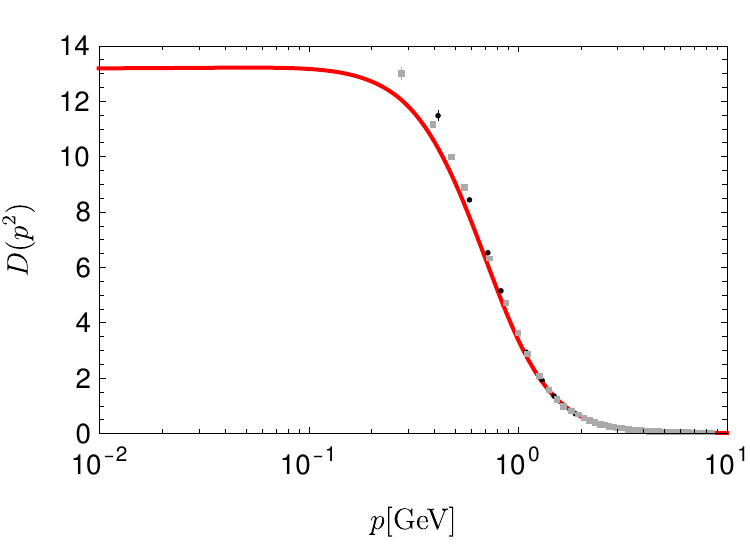}
	\caption{Gluon and ghost dressing functions $Z(p^2)$ and $G(p^2)$, respectively, (left) and gluon propagator $D(p^2)$ (right) in comparison to lattice data \cite{Sternbeck:2006rd}.
	}
	\label{fig:props}
	\includegraphics[width=0.48\textwidth]{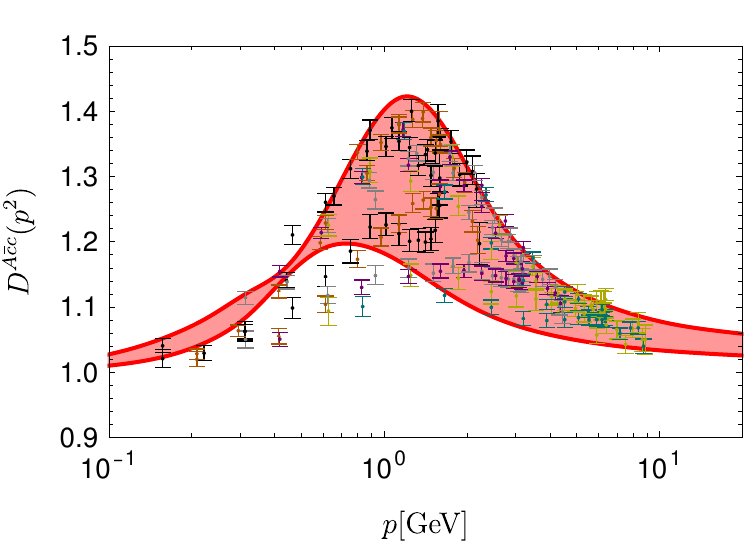}\hfill
	\includegraphics[width=0.48\textwidth]{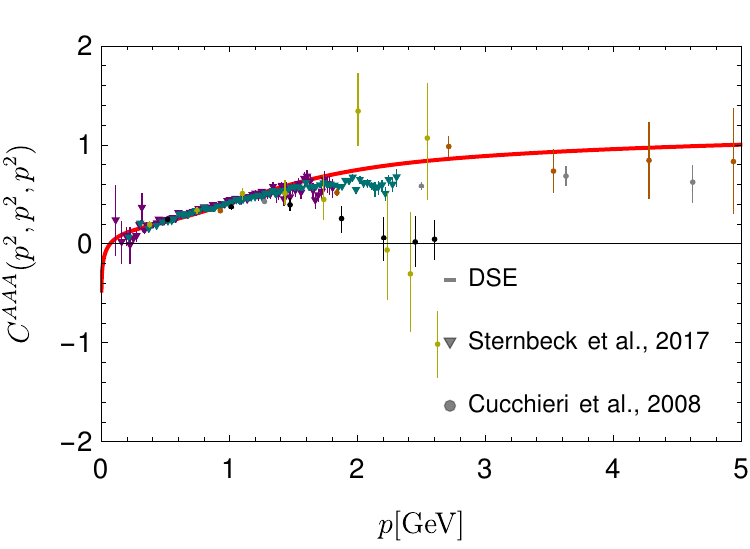}
	\caption{
		Left: Ghost-gluon vertex dressing function (full kinematic dependence) in comparison to $SU(2)$ lattice data \cite{Maas:2019ggf}.
		Right: Three-gluon vertex dressing function at the symmetric point in comparison to lattice data \cite{Cucchieri:2008qm,Sternbeck:2017ntv}), see Refs.~\cite{Athenodorou:2016oyh,Boucaud:2017obn} for similar results.
	}
	\label{fig:verts}
\end{figure*}

We use the Bethe-Salpeter equation (BSE) for a two-gluon bound state as derived from the 3PI effective action truncated to three loops \cite{Berges:2004pu,Carrington:2010qq} as described in \cite{Huber:2020ngt}.
Details on the method can be found in \cite{Fukuda:1987su,McKay:1989rk,Sanchis-Alepuz:2015tha}.
This yields the equations and the corresponding kernels depicted in Figs.~\ref{fig:bses} and \ref{fig:kernels}, respectively.

As input, two and three-point functions of gluons and ghosts are required.
We take them from Ref.~\cite{Huber:2020keu} where they were calculated from the same action.
Note that in this setup no model parameters exist.
There is only one parameter which is the physical scale.
We set the scale by comparing the gluon propagator with corresponding lattice results.
For the comparison with the lattice glueball results, we finally adjust the scale to the same value of the Sommer parameter $r_0=0.472\,\text{fm}$ taken from \cite{Athenodorou:2020ani}.
Figs.~\ref{fig:props} and \ref{fig:verts} show the employed propagator and vertex results together with lattice results.
We want to stress that we do not have to rely only on the good agreement of the functional with the lattice results as an indicator of the reliability of the input, but there are additional tests that attest to its quality.
Obvious checks of a truncation are assessments of the effects of extensions.
Here, we have several such extensions there were performed and found to be quantitatively subleading, among them the inclusion of further four-point functions \cite{Huber:2016tvc,Huber:2017txg} and the use of the full three-gluon vertex basis \cite{Eichmann:2014xya}, see \cite{Huber:2018ned} for an overview.
Last but not least, the agreement with results from the functional renormalization group \cite{Cyrol:2016tym} is also a nontrivial test, as it provides a completely independent set of equations.
Similar extensions of the truncation can be done for the BSE which we discuss in a moment.

It should be noted that several solutions can be realized that differ in the infrared, see, e.g., \cite{Boucaud:2008ji,Fischer:2008uz,Alkofer:2008jy,Maas:2009se,Maas:2011se,Sternbeck:2012mf,Huber:2018ned,Eichmann:2021zuv}.
However, physically these solutions seem to be equivalent and they might be related \cite{Fischer:2008uz,Maas:2009se} to the Gribov problem \cite{Gribov:1977wm,Singer:1978dk,Vandersickel:2012tz}.
We tested that explicitly for the scalar and pseudoscalar glueballs where we found that the masses obtained from different solutions agree within errors \cite{Huber:2021yfy}.
We thus continue here with one solution.

\begin{figure}[tb]
	\includegraphics[width=0.49\textwidth]{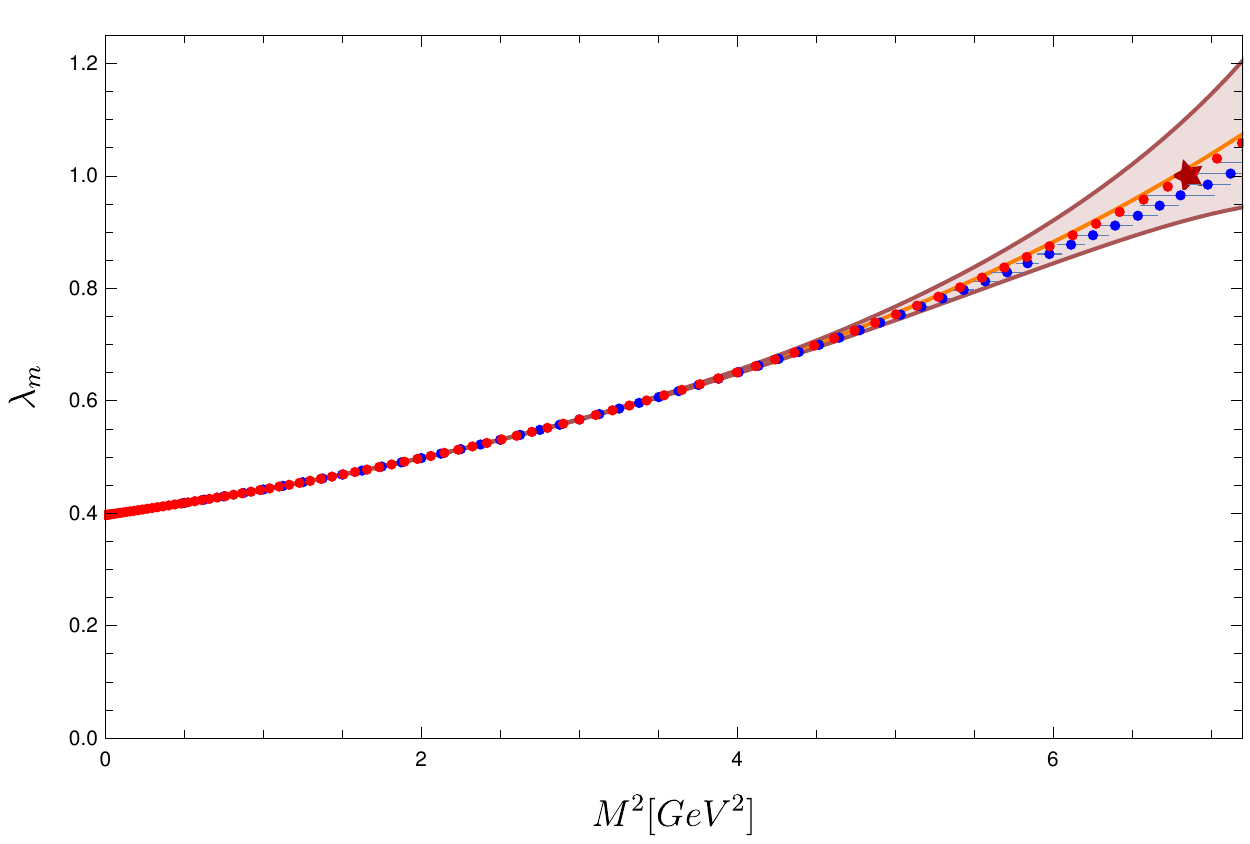}
	\caption{Extrapolation exemplified with a meson of mass $M=2.62\,\text{GeV}$.
        The red dots represent the exact solutions for the eigenvalues, the star the physical one.
		The orange line is the averaged extrapolation with errors indicated by the band.
		The horizontal error bars represent the errors for specific mass values.
		Below $2\,\text{GeV}$, the agreement is so good that the points lie on top of each other.}
	\label{fig:meson_extrap}
\end{figure}

The full BSE splits into two parts which we call glueball-part and ghostball-part, see \fref{fig:bses}.
The respective kernels derived from the 3PI effective action truncated at three loops are shown in \fref{fig:kernels}.
In Refs.~\cite{Huber:2020ngt,Huber:2021yfy}, the diagrams in the red boxes were included which lead to one-loop expressions for the BSE.
For a self-consistent solution of the BSE, the two-loop diagrams are also required.
They are computationally much more expensive, which is why we calculate them with reduced precision.
We checked that this does not affect the ground and first excited states but it can affect the second excited state.
Their inclusion is easiest for the pseudoscalar glueball, because it does not contain a ghostball-part (there is no corresponding amplitude with negative parity).
As it turned out, the two-loop diagrams are completely subleading in this case \cite{Huber:2021zqk}.
There is no significant effect on the masses, as the eigenvalues change by less than 0.1\,\textperthousand.
Here we continue this study for the scalar glueball.
We include the full gluon-gluon interaction kernel.
Since the glueball-part is dominant for the determination of the scalar glueball mass, the resulting two-loop diagrams are expected to yield the largest correction to the original calculation.
For the other interaction kernels, their one-loop expressions are used.

The BSE is solved as an eigenvalue equation for the Bethe-Salpeter amplitude $\Gamma(P,p)$ which depends on the total and relative momenta of the constituents, $P$ and $p$, respectively.
A glueball mass $M^2=-P^2$ is found when the eigenvalue $\lambda(P^2)$ equals one.
The lowest mass corresponds to the ground state and higher ones to excited states.
To solve the equation for time-like momenta $P$, the input needs to be known in the complex plane.
In our case, the input is only available for Euclidean momenta.
Corresponding direct calculations of correlation functions only exist for less advanced truncations
\cite{Strauss:2012dg,Fischer:2020xnb,Horak:2021pfr}.
Instead of extrapolating the input into the complex momentum plane, we solve the BSE for real and positive $P^2$ and then extrapolate the resulting eigenvalue curves to time-like $P^2$.
To this end, we use Schlessinger's continued fraction method \cite{Schlessinger:1968spm,Tripolt:2018xeo}.
To assess its reliability, we discussed a test case that can be solved for time-like $P^2$ in \cite{Huber:2020ngt}.
Up to $2\,\text{GeV}$, the extrapolation is extremely reliable as can be seen in \fref{fig:meson_extrap}.
Beyond that, deviations are observed in the test case.
Their size is estimated by sampling over several different extrapolations, see \cite{Huber:2020ngt} for details.

\section{Results}
\label{sec:results}

\begin{figure*}[tb]
\includegraphics[width=0.48\textwidth]{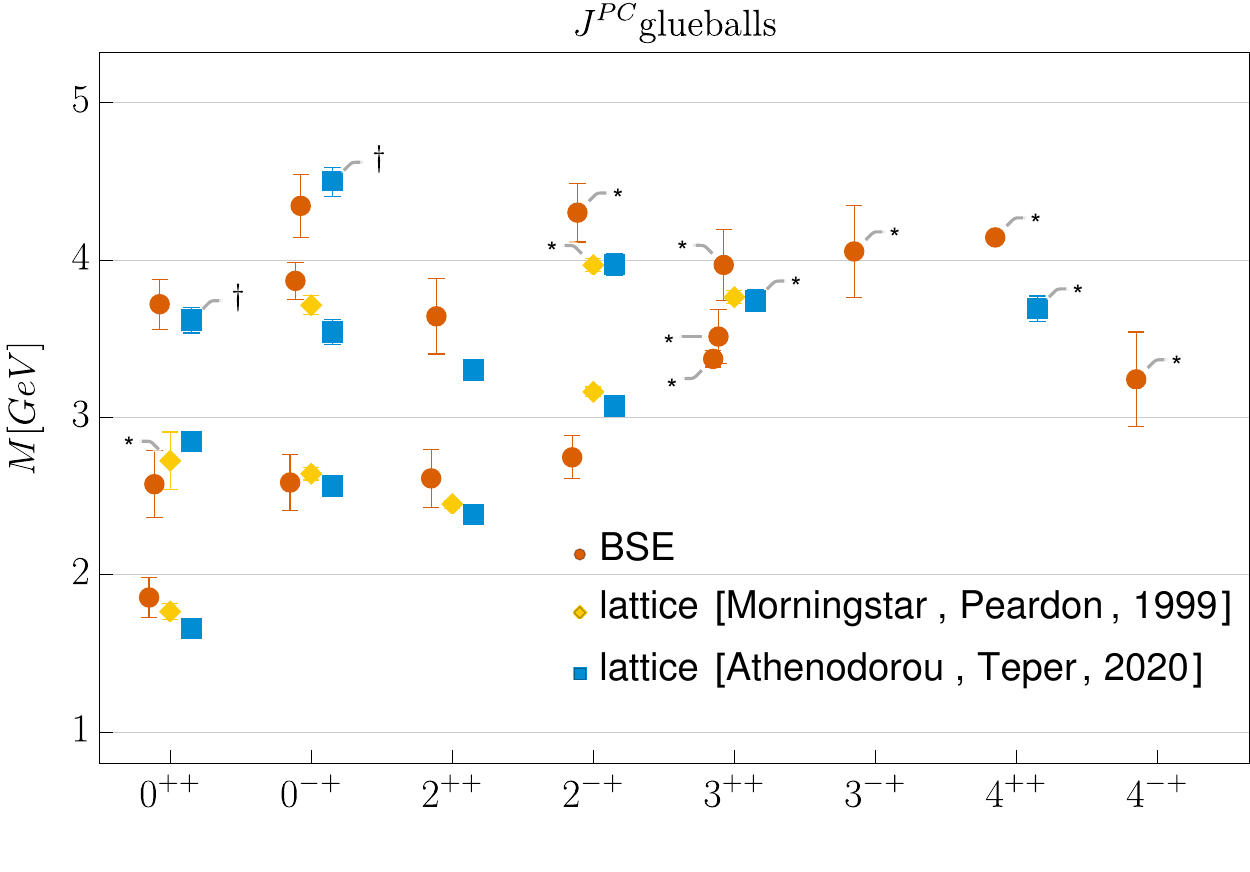}
\hfill
\includegraphics[width=0.48\textwidth]{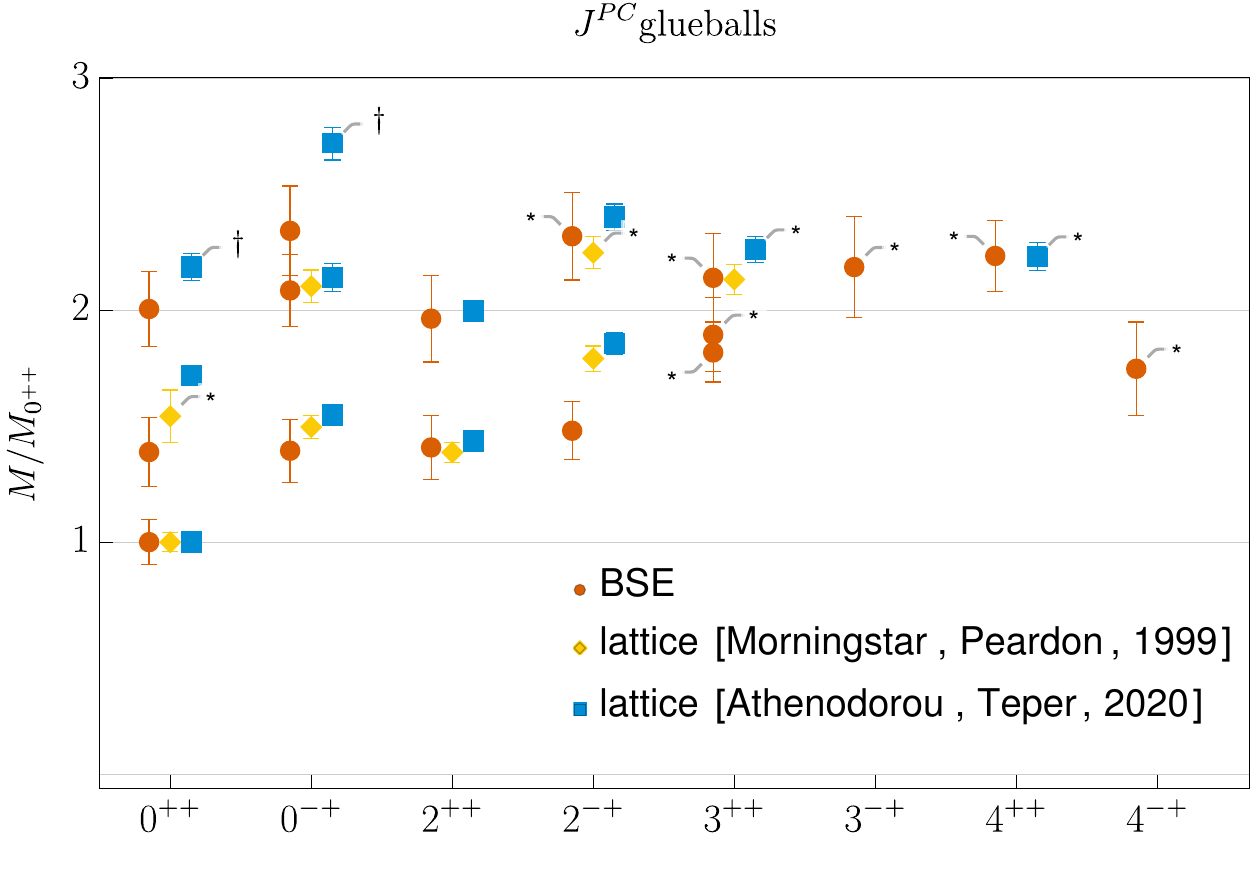}

	\caption{
		Results for glueball ground states and excited states for the indicated quantum numbers from lattice simulations \cite{Morningstar:1999rf,Athenodorou:2020ani} and functional equations.
		In the left plot, we display the glueball masses on an absolute scale set by $r_0=1/(418(5)\,\text{MeV})$.
	    In the right plot, we display the spectrum relative to the ground state.
	    Masses with $^\dagger$ are conjectured to be the second excited states.
		Masses with $^*$ come with some uncertainty in their identification in the lattice case or in the trustworthiness of the extrapolated value in the BSE case.
		}
	\label{fig:spectrum}
\end{figure*}

The results for the quantum numbers $J^\mathsf{CP}=0^{\pm+},2^{\pm+},3^{\pm+},4^{\pm+}$ are shown in \fref{fig:spectrum} and \tref{tab:masses}.
We also solved the BSE for spin $J=1$ but did not obtain sensible solutions.
We want to stress that this is a consequence of the dynamics of the two-body equation and not of the Landau-Yang theorem \cite{Landau:1948kw,Yang:1950rg} which does not apply in this framework, because gluons are not on-shell \cite{Huber:2021yfy}.
For the lighter states, there is good agreement with lattice results.
In some cases, we find even second excited states.
For heavier states, the uncertainty due to the extrapolation increases.
We denote this uncertainty in the plots and the table by $^*$.
Also, heavy states can decay which is, however, not captured by the employed truncation.

For the scalar glueball, we can compare different levels of truncation.
Originally, all one-loop diagrams were included \cite{Huber:2020ngt}.
Here we compare that to a calculation where the two-loop diagrams for the glueball-part are included.
Although this has only a very small effect (less than a per mille) on the individual eigenvalues, their extrapolation towards the physical mass is shifted slightly.
These shifts, however, are still much smaller than the total extrapolation error.
For the ground state, the mass is one percent higher, for the first excited state two percent.
For the second excited state we could not make a comparison.
The reason is that for the two-loop calculation we had to reduce the numeric precision which, as we tested explicitly, is harmless for the first two states but not the second excited state.
We thus conclude that the two-loop diagrams from the gluon-gluon interaction kernel are severely suppressed.
The neglected two-loop diagrams from the ghost-gluon and ghost-ghost interaction kernels are expected to be even more irrelevant as the ghostball-part itself is subleading.
Neglecting it decreases the mass only by approximately three percent for the groundstate and fourteen percent for the first excited state.

\begin{table*}[tb]
\begin{center}
		\begin{tabular}{|l||c|c|c|c|c|c|c|c|}
			\hline
			&  \multicolumn{2}{c|}{\cite{Morningstar:1999rf}} & \multicolumn{2}{c|}{\cite{Chen:2005mg}} & \multicolumn{2}{c|}{\cite{Athenodorou:2020ani}} & \multicolumn{2}{c|}{This work}\\   
			\hline
			State &  $M\, [\text{MeV}]$& $M/M_{0^{++}}$ & $M\, [\text{MeV}]$& $M/M_{0^{++}}$  &  $M\, [\text{MeV}]$& $M/M_{0^{++}}$ & $M\,[\text{MeV}]$ & $M/M_{0^{++}}$\\   
			\hline\hline
			$0^{++}$ & $1760 (50)$ & $1(0.04)$ & $1740(60)$ & $1(0.05)$ & $1651(23)$ & $1(0.02)$ & $1850 (130)$ & $1(0.1)$\\
			\hline
			$0^{^*++}$ & $2720 (180)^*$ & $1.54(0.11)^*$ & -- & -- & $2840(40)$ & $1.72(0.034)$ & $2570 (210)$ & $1.39(0.15)$\\
			\hline
			\multirow{2}{*}{$0^{^{**}++}$} & \multirow{2}{*}{--} & \multirow{2}{*}{--} & \multirow{2}{*}{--} & \multirow{2}{*}{--} & $3650(60)^\dagger$ & $2.21(0.05)^\dagger$ & \multirow{2}{*}{$3720 (160)$} & \multirow{2}{*}{$2.01(0.16)$}\\
			& & & & & $3580(150)^\dagger$ & $2.17(0.1)^\dagger$ & &\\
			\hline
			$0^{-+}$ & $2640 (40) $ & $1.50(0.05)$ & $2610(50)$ & $1.50(0.06)$ & $2600(40)$ & $1.574(0.032)$ & $2580 (180)$ & $1.39(0.14)$\\
			\hline
			$0^{^*-+}$ & $3710 (60)$ & $2.10(0.07)$ & -- & -- & $3540(80)$ & $2.14(0.06)$ & $3870 (120)$ & $2.09(0.16)$\\
			\hline
			\multirow{2}{*}{$0^{^{**}-+}$} & \multirow{2}{*}{--} & \multirow{2}{*}{--} & \multirow{2}{*}{--} & \multirow{2}{*}{--} & $4450(140)^\dagger$ & $2.7(0.09)^\dagger$ & \multirow{2}{*}{$4340 (200)$} & \multirow{2}{*}{$2.34(0.19)$}\\
			& & & & & $4540(120)^\dagger$ & $2.75(0.08)^\dagger$ & &\\
			\hline
			\hline
			$2^{++}$ & 2447(25) & 1.39(0.04) & 2440(50) & 1.40(0.06) & 2376(32) & 1.439(0.028) & 2610(180) & 1.41(0.14)\\
			\hline
			$2^{^*++}$ & - & - & - & - & 3300(50) & 2(0.04) & 3640(240) & 1.96(0.19)\\
			\hline
			$2^{-+}$ & 3160(31) & 1.79(0.05) & 3100(60) & 1.78(0.07) & 3070(60) & 1.86(0.04) & 2740(140) & 1.48(0.13)\\
			\hline
			$2^{^*-+}$ &  3970(40)$^*$ & 2.25(0.07)$^*$ & - & - & 3970(70) & 2.4(0.05) & 4300(190) & 2.32(0.19)\\
			\hline\hline
			$3^{++}$ & 3760(40) & 2.13(0.07) & 3740(60) & 2.15(0.09) & 3740(70)$^*$ & 2.27(0.05)$^*$ & 3370(50)$^*$ & 1.82(0.13)$^*$\\
			\hline
			$3^{^*++}$ & - & - & - & - & - & - & 3510(170)$^*$ & 1.89(0.16)$^*$\\
			\hline
			$3^{^{**}++}$ & - & - & - & - & - && 3970(220)$^*$ & 2.14(0.19)$^*$\\
			\hline
			$3^{-+}$ & - & - & - & - & - & - & 4050(290)$^*$ & 2.19(0.22)$^*$\\
			\hline
			\hline
			$4^{++}$ & - & - & - & - & 3690(80)$^*$ & 2.24(0.06)$^*$ & 4140(30)$^*$ & 2.23(0.15)$^*$\\
			\hline
			$4^{-+}$ & - & - & - & - & - & - & 3240(300)$^*$ & 1.75(0.2)$^*$\\
			\hline
		\end{tabular}
		\caption{Ground and excited state masses $M$ of glueballs for various quantum numbers.
		Compared are lattice results from \cite{Morningstar:1999rf,Chen:2005mg,Athenodorou:2020ani} with the functional results of \cite{Huber:2020ngt,Huber:2021yfy}.
        For \cite{Morningstar:1999rf,Chen:2005mg}, the errors are the combined errors from statistics and the use of an anisotropic lattice.
		For \cite{Athenodorou:2020ani}, the error is statistical only.
		In our results, the error comes from the extrapolation method and should be considered a lower bound on errors.
		All results use the same value for $r_0=1/(418(5)\,\text{MeV})$.
		The related error is not included in the table.
		Masses with $^\dagger$ are conjectured to be the second excited states.
		Masses with $^*$ come with some uncertainty in their identification in the lattice case or in the trustworthiness of the extrapolated value in the BSE case.
		}
		\label{tab:masses}
	\end{center}
\end{table*}

\FloatBarrier

\section{Summary}
\label{sec:summary}

We presented results for the glueball spectrum of quenched QCD calculated from the 3PI effective action.
We obtain good agreement with lattice results where they are available and add more states.
For the scalar glueball, we extended the calculations from \cite{Huber:2020ngt,Huber:2021yfy} by including two-loop diagrams.
As opposed to the pseudoscalar glueball, where these diagrams are totally negligible \cite{Huber:2021zqk}, we do find a nonzero effect for the scalar glueball.
However, it is smaller than the extrapolation error.
This provides further evidence that the employed truncation is quantitatively reliable.
For the inclusion of quarks, as planned in future work, a similar truncation is thus promising.

\section*{Acknowledgments}

This work was supported by the DFG (German Research Foundation) grant FI 970/11-1 and by the BMBF under 
contracts No. 05P18RGFP1 and 05P21RGFP3.
This work was also supported by Silicon Austria Labs (SAL), owned by the Republic of Austria, the Styrian Business Promotion Agency (SFG), the federal state of Carinthia, the Upper Austrian Research (UAR), and the Austrian Asso­ci­a­tion for the Elec­tric and Elec­tronics Industry (FEEI).

\bibliographystyle{utphys_mod}
\bibliography{literature_glueballs_hadrons21}

\end{document}